# Super-Atom Representation of High-T$_C$ Superconductivity

Itai Panas

Environmental Inorganic Chemistry

Department of Chemistry and Biotechnology

Chalmers University of Technology

S-412 96 Gothenburg, Sweden

**Abstract**

A resonating valence bond RVB approach is taken to demonstrate formation of real-space Cooper pairs and High-T$_C$ superconductivity HTS. Non-adiabatic coupling between holes aggregates (super-atoms) and undoped anti-ferromagnet cause virtual excitations in either system due to inter-system coupling. HTS is said to reflect cooperative co-existence of two Bose-Einstein condensates in terms of one real-space Cooper pair condensate, and a second magnon condensate, which form at the same critical temperature. T$_C$ is formulated in terms of the super-exchange interaction. Connection is made to an equivalent real-space BCS formulation of HTS. Novel perspectives on the HTS in the electron-doped $Sr_{1-x}La_xCuO_2$ and $Nd_{2-x}Ce_xCuO_4$ emerge.

## 1. Introduction

A quarter of a century has passed since the promising discovery of high critical temperature superconductivity HTS in the cuprates [1], and still no consensus on the decisive mechanism for the phenomenon is at hand. Many theories have been proposed to come to terms with the HTS, some focusing on the phase diagram [2], others building on quantum criticality [3,4], and yet others base their understanding on an effective single-band scenario into which holes are doped [5, 6] and conditions discussed for HTS to emerge. All these theories are unique to the cuprates in that they carry little or no predictive value in the search for new HTS materials. From the perspective of the theoretical chemist, it becomes important to provide a chemically simple yet sufficiently complex conceptual model [7-9] to leave room for chemical manipulations as well as predictive quantum chemical model calculations [10-13]. The message to the physics community is in the form, which a viable mechanistic understanding of HTS must take in order to be acceptable from a chemical perspective.

During the last decade the predicted checkerboard structure [7,12,14-17], presence and essential roles of short-range AFM fluctuations [18] and in particular the magnetic spin-flip excitation [7,12,19-21], c-axis dynamics of the ions at the A-sites [7,9,10,22], D-wave order parameter symmetry [7,23] and a fine local energy scale in vortex cores [7-13,24] have all been confirmed by experiment as being essential. In particular, these observations are all consistent with an understanding, which has the HTS develop from a pseudo-gapped state [25], and contradict scenarios based on HTS emerging from well developed Fermi surfaces [26,27]. With the recent advent of superconductivity in the Fe-chalcogenide [28] and Fe-pnictide [29] materials, there is a growing impression in the solid state community of the existence of a possible general underlying principle for achieving SC in the so-called strongly correlated systems, where the cuprates and the iron based materials are but two representatives of the class. In this context a common such possible understanding was formulated in [30].

Today it is acknowledged that the Bardeen Cooper Schrieffer BCS theory for superconductivity [31] is but the simplest possible formulation, as it succeeds in demonstrating how an electron gas model system may undergo transition into a superconductor. However, taking the electron-gas hamiltonian as 0:th order ansatz is in general not physical, and particularly not so for the so-called strongly correlated systems, such as the cuprates and the new iron based superconductors. It was the identification of the SC phenomenon with its BCS theory representation which rendered particular drama to the discovery of HTS in the strongly correlated systems. Materials which display strong correlations were thought to offer all the reasons for not turning superconducting, such as spin-density waves, charge-density waves, giant magneto-resistance, colossal magneto resistance, orbital ordering, ferromagnetic metallicity, antiferromagnetism etc. These properties are all signatures of instabilities, which are realized by corresponding symmetry breakings. Still, the highest critical temperatures for superconductivity were found in the strongly correlated systems. In [30] it was proposed that what distinguishes the superconductors among the strongly correlated systems is that when all the possible symmetry breakings have occurred, of which short-range AF order is essential, in some systems local electronic near-degeneracy still remains to be lifted by the development of a pseudogaped Fermi surface. The pseudogap is proposed to reflect instability of local origin, e.g. in terms of holes clustering (hole-doped cuprates) or among local d-states (Fe-superconductors). The non-adiabaticity between short-range AF and local charge carrier degrees of freedom was said to provide means for the system to lower its energy as this coupling allows for improved electron correlation, and the consequence of which is the non-local phase rigidity of the SC.

The objective of the present study is to build on the above understanding and possible confirmations, in order to provide a conceptual tool for the discovery of new high-$T_C$ superconductors. It is recognized that the emerging understanding in [7,30] shares important common flavors with the Resonating Valence Bond RVB model of Anderson [32] and claimed relevance to HTS [33, 34]. The purpose here is to articulate in an RVB context the possible understanding summarized in [7,30], and to demonstrate how the super-atom representation can be generalized to express superconductivity in the so-called electron doped $Nd_{2-x}Ce_xCuO_4$ superconductor.

## 2.    Conceptual understanding

Essential characteristics of the Anderson RVB theory comprise spin-charge separation in a doped Mott-Hubbard insulator. In that representation, because the charge excitations do not carry spin, they may undergo Bose-Einstein condensation. A key feature of that theory is almost-free holes amplitudes displaying the symmetry of bound electron pairs. In parallel, the quantum chemistry based conceptual model achieves said "separation" by the charge-carriers physically separating from the AFM band, thus forming hole clusters, i.e. "super-atoms", and it has HTS emerging from a *resonance requirement* for excitations in the anti-ferromagnet with excitations in the super-atoms. The RVB understanding of holes-pairs amplitudes being complementary to the "bound electrons" such that the former may under-go BEC, is realized in detail in the super-atoms representation (see below), where virtual excitations in the AFM are required to produce the coupling between super-atoms.

The chemical perspective is based on local electronic motifs, i.e. "super-atoms" and states that superconductivity emerges upon achieving electronic phase coherence among such objects. In [7-13] charge carriers clustering was discussed and demonstrated to occur as a response to disorder in the position of the ions at the A-site. Let these hole clusters (see Figure 1) constitute said "super-atoms" and interpret the sharing of a pair-state among precisely two such "super-atoms" to reflect the formation of a real-space Cooper pair.

Fig.1

Such an ansatz is justified based on the locality constraints of non-adiabatic coupling of two electronic sub-systems where one is propagated by superexchange interactions in conjunction with the assumption that HTS is a general phenomenon. Note in e.g. Figure 1b how the AFM band characteristics, seen best in the spin density on the rim of the 4 x 4 super-cell, comes out different from the "super-atom" spin density in the center. This observation illustrates the cause for our dual formulation of superconductivity, and our deviation from the RVB understanding in its simplest form. Hence, it deviates from the charge-spin separation in a single band in that charge carriers are allowed to leave the AFM band altogether (see Figure 1 again), and build local super-atoms in terms of states that belong to a low-dispersive second band (See Figure 2, Top: AFM states; Bottom: "Super-atom" states).

Fig.2

Thus, the set of non-bonding oxygen orbitals acts as buffer for holes. Yet, our approach claims to share essential characteristics with RVB, and can be understood as a generalized RVB. This is explicitly demonstrated below.

The simplest way to express the real-space Cooper pair employs two local wave functions $\Psi_1$ and $\Psi_2$ to represent local super-atoms at site 1 and site 2, where four such sites are shown in Figure 1. Let the ground state of the super-atom be a node-less pair-state, which implies that the first excited state is pair-broken and orthogonal to the pair state, as reflected e.g. in the deviating node structure. We postulate that entering into the superconducting state is equivalent to writing the local ground states at *site i* as a superposition of super-atom pair- (P) and pair-broken (PB) states

$$\Psi_i = u_i \cdot PB_i + v_i \cdot P_i \tag{1}$$

This wave function ansatz was exemplified and discussed in some depth in [7]. It implies a local "super-symmetry", such that super-atom pair- and pair-broken states mix. This is made possible by the coexistence of a second electronic subsystem, which displays local antiferromagnetic order, and acts as spin and symmetry buffer by displaying the complementary local virtual magnon to match the virtual super-atom excitations (see Figure 2, [7]). Thus, neither super-atom nor local AFM channels preserve local spin and space symmetry. Instead local spin and space symmetry is preserved locally for the local

AFM⊗Super-atom compound. Access to this effective super-symmetry for the super-atoms channel comes at the expense of lifting "non-locality symmetry", as it requires phase coherence among the super-atoms. Similarly, the required complementary virtual magnons in the local AFM are understood as lifting the short-range magnetic order. The local "super-symmetry" implied by appearance of virtual magnons in a local antiferromgnet at 0 K, comes at the expense of detailed coupling in the super-atoms channel. The two corresponding so-called Nambu-Goldstone bosons are reflected in (a) the quasi-particle excitation out of the superconducting condensate and (b) the magnetic spin-flip excitation [19-21]. The latter tells of phase rigidity in the magnetic channel enforced by the superconductivity. It is reflected in magnetic super-symmetry, i.e. mixing of virtual local singlet-triplet states in the magnetic channel, complementary to the corresponding virtual excitations in the super-atoms.

## 3. RVB among Super-atoms

In order to realize essential aspects of this understanding, assume that the super-atom wave functions (see Eq. 1) are individually normalized and that inter-site overlaps vanish. This excludes coherence being maintained by overlap or tunneling. Moreover, let $\langle PB_i | PB_i \rangle = 1, \langle P_i | P_i \rangle = 1$, and $\langle P_i | PB_i \rangle = 0$, and assign the energy $\alpha_i$ for each super-atom, such that

$$H_{ii} = \langle \Psi_i | H | \Psi_i \rangle = \alpha_i \qquad (2)$$

Let pair-wise interactions between super-atoms delocalize the singlet pair-state according to the coupling matrix element

$$\langle H_{12}\rangle = \langle \Psi_1\Psi_2 | C^+_{2,P}C_{2,PB_1} <\Psi_M | M^+_{2,P}M_{2,PB_{-1}} |W| M^+_{1,PB_{-1}}M_{1,P} | \Psi_M> C^+_{1,PB_1}C_{1,P} | \Psi_1\Psi_2\rangle \quad (3)$$

where $C_{i,\Sigma_S}$ ($M_{i,\Sigma_S}$) annihilates a local state $\Sigma$ (P/PB in superatom, and AF/magnon in the AFM) with projected spin ($S_Z=0,\pm 1$) in the super-atom (AFM) and $C^+_{i,\Sigma_S}$ ($M^+_{i,\Sigma_S}$) creates corresponding local states (see e.g. Figure 2). Eq. 3 describes how a local transition from the pair-state into the pair-broken state in one super-atom, is accompanied by the complementary transition from a pair-state into a pair-broken state in the magnetic channel in the vicinity of that site, while preserving the local compound electronic spin and space symmetry. This is accompanied in Eq. 3 by the precise complementary process at a second site. Introducing Eq. 1 into Eq.3 we obtain

$$H_{12} = v_1 \cdot u_1 \cdot u_2 \cdot v_2 \cdot J_{local} = \beta \quad (4)$$

where $J_{local}$ is a measure of the local magnon energy in the vicinity of each cluster. This is an attractive interaction because only by such coherent inter-cluster coupling does $J_{local}$ survive the holes doping [12]. This lends crucial significance to the magnetic spin-flip resonance in inelastic neutron scattering [19, 20] observed at $T \leq T_C$ and consistent with recent μ-spin resonance measurements [21]. Let

$$u_1 = u_2 = u$$

$$v_1 = v_2 = v$$

$$\alpha_i = \alpha$$

and solve the Schrödinger equation for the wave function ansatz

$$\Phi_\pm = c_1\Psi_1 \pm c_2\Psi_2 \quad (5)$$

The resulting secular equation implies solving for

$$\begin{vmatrix} \alpha - E & \beta \\ \beta & \alpha - E \end{vmatrix} = 0 \qquad (6)$$

Eigenvalues and eigenstates come out as

$$E = \alpha \pm \beta \text{ and } c_1 = c_2 = \frac{1}{\sqrt{2}} \qquad (7)$$

In particular, the energy splitting between the two states

$$E_2 - E_1 = \Delta E = 2\beta = 2 \cdot u^2 v^2 J_{local} \qquad (8)$$

determines the stability of the delocalized pair-state (real-space Cooper pair) shared by the interacting super-atoms Equivalently, let $n_s$ be the super-fluid pair-density. Then we have

$$\Delta E \propto 2 \cdot n_s J_{local} \qquad (9)$$

The energy gap (Eq. 9) is a local analog to the BCS gap, i.e. local states in the interval $E_1 < E < E_2$ across the Fermi level are expelled due to the inter-site coupling (3).

Write $\Delta E$ (Eq. 8) in terms of local pair- and pair-broken state probability densities

$$n_P = v^2, \quad n_{PB} = u^2 \qquad (10)$$

such that

$$\Delta E = 2 n_P n_{PB} J_{local} \qquad (11)$$

where we have $n_P n_{PB} \propto n_s$. Trivial identities which emerge are

$$\Delta E = 2 \cdot n_P (1 - n_P) J_{local} = 2 \cdot n_{PB} (1 - n_{PB}) J_{local} \propto 2 n_s J_{local} \qquad (12)$$

If such shared pair-states constitute the very bosons, which subsequently are to form the superconducting Bose-Einstein condensate and the BEC is limited by the stability of the real-space Cooper pair, then we must have $\Delta E \propto T_C$, i.e. the stability of this delocalized pair-state determines the critical temperature

$$T_C \propto n_P n_{PB} J_{local} \propto n_s J_{local} \tag{13}$$

### 4. Phase diagram – Hole doping *versus* Temperature

The ability to represent local co-existence of pair- and pair-broken super-atom states has implications on the $T_C$ versus doped holes phase diagram of the cuprates (*cf.* analysis in [21]). In case of the hole doped cuprates for a 4-holes super-atom, we have rapid suppression of the $T_C$ when doping beyond 4 holes per 16 $Cu^{2+}$ sites (See Figure 1). Generalize (Eq. 13) to say that

$$T_C(x) \propto n_P(x) \cdot n_{PB}(x) \cdot J_{local} \propto n_s(x) \cdot J_{local} \tag{14a}$$

where $x$ is the hole concentration and $n_P(x)$, $n_{PB}(x)$, and $n_s(x)$, are the doping dependent analogs to the doping independent $n_P$, $n_{PB}$, and $n_s$, defined above. Assume that the doping dependences of said probability densities are associated with the probability of finding a hole cluster. Two extremes may be considered:

<u>Either</u> the super-atom is the only charge carrier inhomogeneity, in case of which we get

$$n_P(x) \propto (1-4x) \cdot n_P,$$

$$n_{PB}(x) \propto 4x \cdot n_{PB}$$

$$n_s(x) \propto n_P(x) \cdot n_{PB}(x) \propto 4x \cdot (1-4x) \cdot n_s$$

$$T_C \propto n_s \cdot 4x \cdot (1-4x) \cdot J_{local} \tag{14b}$$

<u>Or</u> the superconductivity emerges from a competing striped charge density wave phase [35], then for $\frac{1}{8} \leq x \leq \frac{1}{4}$ we obtain

$$T_C \propto n_s \cdot 8(x - \tfrac{1}{8}) \cdot (1 - 8(x - \tfrac{1}{8})) \cdot J_{local} \qquad (14c)$$

## 4. Local Antiferromagnet Perspective - Consequences of Complementarity

It is implied above that achieving superconductivity by magnon mediated Bose-Einstein condensation of resonating super-atom pair-states, i.e. real-space Cooper pairs, necessarily means corresponding BE condensation of virtual magnons. Let $m$ be the virtual magnon probability density, and $2\delta_{local}$ the local cluster pair-breaking excitation. Taking this perspective to be complementary to the above, we write (*cf.* Eqs. 11, 12):

$$\Delta E = 2 \cdot m(1-m)\delta_{local} \qquad (15a)$$

and analogously to Eq. 9

$$\Delta E = 2 \cdot n_m \delta_{local} \qquad (15b)$$

where $m(1-m) \propto n_m$, and $n_m$ is the BE condensate of virtual magnons. Because BE condensation in the super-atoms channel occurs simultaneously with that in the AFM, we equate the two expressions for $\Delta E$ to obtain

$$\frac{J_{local}}{\delta_{local}} = \frac{n_m}{n_s} = \frac{m}{n_{PB}} \cdot \frac{1-m}{n_P} \qquad (16)$$

Let $\delta_{local} = J_{local}$ be the resonance requirement for the applicability Eq. 3, or equivalently that $n_s = n_m$. This implies that for every $\delta_{local}$ there is a $J_{local}$ up to $J_{max}$ which corresponds to the superexchange coupling in the undoped system. The origin of distributions of $\delta_{local}$ can be understood as a disorder in the A-site, which affects $J_{local}$ due to bond lengths changes in adjacent in-plane Cu-O bonds [22]. The two solutions of Eq.16 are:

(i) $m = n_{PB}$, i.e. virtual magnon density equals virtual cluster pair-breaking density; or,

(ii) $m = n_P$, i.e. virtual magnon density equals virtual cluster pair-density.

At $\Delta E_{max}$ the two are equivalent. Thus, an upper bound to the $T_C$ can be estimated from

$$\Delta E_{max} = \frac{1}{2}\delta_{local} = \frac{1}{2}J_{local} \qquad (17)$$

For $J_{local} = J_{max}$, and assuming $J_{max}$ to reflect the Neél temperature in the same way as $\Delta E_{max}$ does $T_C$, we obtain

$$T_C^{max} \sim \frac{1}{2}T_N \qquad (18)$$

Hence, we have employed the equivalence of two complementary descriptions of the superconductor, i.e. taking the "super-atoms" perspective and that of the antiferromagnet, respectively, in order to illustrate the close relation between the Neél temperature of the undoped system, and the maximum critical temperature for superconductivity. The obtained estimate is in line with recent reports on possible local superconducting correlations above $T_C$ in the cuprates [36].

5.     **Making contact between RVB-BEC and Real-Space BCS**

For completeness, we make connection between the BEC discussion here and the real-space BCS interpretation in [7]. For $V$, said to be the energy gained from charge carrier pairing and associated to the resulting recovery of $J_{local,}$, the expression

$$\Delta^2(0K) = \left(\frac{x}{2}\right)^2 V^2(\mu+\nu)^2 - \gamma^2(\mu-\upsilon)^2 \qquad (19)$$

was obtained for the superconducting gap at T=0 K; for definitions of $\mu,\upsilon,\gamma$ see [7]. For x=1, superconductivity emerges from pair-wise cluster-cluster interactions accommodating the delocalized pair-states. This situation corresponds to the above formations of real-space Cooper pairs. Equating $\Delta E_{max}$ with $2\Delta_{max}(0K)$, we get for $\mu = \upsilon = \frac{1}{\sqrt{2}}$

$$2\Delta_{max}(0K) = \sqrt{2} \cdot V = \frac{1}{2} J_{max} \qquad (20)$$

In the present study we emphasize the BEC aspect of the HTS phenomenon. It articulates the co-existence of two condensates, one in the AFM excitations and a second in the super-atoms excitations, and the one required in order for the other to form, such that the condensations occur at one and the same critical temperature. It is gratifying to note that room-temperature Bose-Einstein condensation of magnons, obtained upon pumping, has recently been reported [37].

The above RVB-BEC exercise in conjunction with the real-space BCS formulation [7] constitutes two consistency checks for our understanding of HTS. This understanding offers two complementary perspectives of one and the same mechanism. The essential elements comprise two *a priori* disjoint electronic sub-systems. Resonant coupling between the two enforces non-local phase coherence, which can be interpreted as the coexistence of two Bose-Einstein condensates, one offering the coupling for the other.

### 6. Implications for the holes doped cuprates

In case of the cuprates, two complementary and equivalent approaches to the HTS phenomenon emerge. One perspective takes the single-band Hubbard model for a Mott insulator as central component. The required spontaneous excitations $0 < J_{local} < J_{max}$ in the AFM are achieved by resonant coupling to corresponding spontaneous excitations between local charge carrier cluster states, made possible by the *inter-cluster delocalized pair-amplitudes* (i.e, real-space Cooper pairs), such that for each $J_{local}$ there is a $\delta_{local}$ set by the local A-site cation-to-plane distance. The second perspective takes the inter-cluster delocalization of pair-states (real space Cooper pairs) as starting point for an effective theory of HTS. Local magnon excitations in the antiferromagnetic background are required in order to preserve local spin and space symmetry as pair-and pair-broken super-atom states are allowed to mix. This mix is precisely the origin of the D-wave superconductivity [7, 23]. For all cluster excitations in the range $0 < \delta_{local} < J_{max}$ there is a resonant virtual magnon. This renders both magnon and cluster excitations virtual processes.

The two perspectives are obviously complementary, and preference for one over the other is decided by particularities in any relevant question at hand. The main achievement here is for synthesis of new high-Tc superconductors, i.e. separation of the HTS into subsystems with detailed properties that may be formulated separately, and where detailed matching of the two subsystems upon assembly produces the superconductivity. Possible recent evidence for such separability is provided in a recent ARPES study of the pseudo-gap state [38], which reports particle-hole asymmetry as well as spatial symmetry breaking. Connection to the understanding presented here , i.e. in terms of two bands and super-atoms formations, is made in [39]. It is concluded that that the pseudo-gapped state observed by Hashimoto et al. [38] is the precursor to the superconducting state and not some competing phase.

## 7. In Search for New HTS Materials – Conclusions

The above understanding has recently been employed to describe the appearance of superconductivity in α-FeSe [30]. Two possible understandings regarding the cuprates emerge. In both cases the common interpretation, i.e. electron doping of the $Cu3d_{x^2-y^2} - O2p_\sigma$ band, is replaced by electrons segregation. One motif implies formation of cationic vacancies. Thus is the infinite layer system [40-41] taken here to imply formation of $Sr_{1-\frac{3}{2}x}La_xCuO_2$, i.e. replacement of three Sr by two La and an Sr vacancy. In this case, crystal field inhomogeneity causes formation of Cu3d$^9$ in conjunction with δCu3d$^{10}$ and O$^{2-\delta}$ followed by holes clustering (see Figure 3). Possible S-wave symmetry has been reported for this system [42], in line with possible absence of local AF background.

Fig. 3

In case of the electron doped Nd$_{2-x}$Ce$_x$CuO$_4$ ($x \approx 0.13$) cuprate [43] superconductors a similar interpretation may be employed, resulting in $Nd_{2-\frac{4}{3}x}Ce_xCuO_4$. However, in case of the latter, direct contact between the Nd$_{2-x}$Ce$_x$CuO$_4$ ($x \approx 0.13$) system and the superatoms in the holes-doped cuprates is achieved by allowing electron the Nd to act electron sink instead of the Cu ions. Hence, the "electron doping" implies formation of local Ce-Nd pairs which interact across the CuO$_2$ plane (Figure 4).

Fig. 4

In this case, the local four-electron system from $4f^3$ $Nd^{3+}$ and $4f^1$ $Ce^{3+}$, and the resulting electronic states constructed from 4f-orbitals parallel the crucial symmetry and electron properties with the hole-cluster "super-atom" (compare Figures 2 and 3). In case of $Nd^{3+}$-$Nd^{3+}$ pairs, this is a 6-electron-system which does not satisfy the requirements for achieving the non-adiabatic coupling to the AFM. Thus, while the hole cluster constitutes the universal motif on the holes doping side, alternative motifs are required upon "electron doping" in order to couple to the AFM band in an equivalent way as does the hole-cluster based super-atom.

In summary, it has been demonstrated how the HTS phenomenon may be understood as the cooperative formation of two Bose Einstein condensates, composed of AFM excitations and super-atom excitations, respectively, which condense at the same critical temperature. This has allowed a novel approach to the SC phenomenon in the electron-doped side of the phase diagram of the cuprates superconductors. It is hoped that the super-atom representation of high-Tc superconductivity will provide a conceptual tool for future chemical explorations.

**Acknowledgement**

Professor Amit Keren is gratefully acknowledged for sharing results, inspiring discussions, and warm hospitality during my visit to his group at the Technion in Haifa, Israel. Inspiring discussions with Professors Tord Claeson, and Øystein Fischer are gratefully acknowledged.

*Computational details*

*The CASTEP [44] program package within the Material Studios framework [45] was utilized and the GGA PBE functional [46] employed. Core electrons were described by ultra-soft pseudo-potentials employing a 300 eV cut-off energy. Results from gamma point calculations are presented through-out.*

Figure Caption

**Figure 1**

(A-C) display DFT Spin densities in $HgBa_{1.87}CuO_4$ for 25% holes doping produced by the introduction of 2 Ba vacancies in a superstructure in 4 × 4 unit cells. (A) and (B) are side views and top views respectively. (C) emphasize the checkerboard structure providing possible connection to STM experiments [14-16].

**Figure 2.**

Three panels represent three different electron configurations, which contribute to the ground state. Within each quartet, four electrons are distributed in four linear combinations of atomic orbitals. Top portions of panels refer to the 4 electrons in the magnetic subsystem, while the bottom portions refer to 4-holes super-atom states. **Left panel:** Ground state occupations of AMF and super-atom. **Center panel:** symmetry of excited state in top portion $XY(X^2-Y^2)$ and $S_z=1$ ; symmetry of excited state in bottom portion: $XY(X^2-Y^2)$ and $S_z=-1$. **Right panel**: symmetry of excited state in top portion $(X^2-Y^2)$ and $S_z=0$ ; symmetry of excited state in bottom portion: $(X^2-Y^2)$ and $S_z=0$.

**Figure 3.**

Spin density at an Sr vacancy in $Sr_{0.888}La_{0.074}CuO_2$ (violette), i.e. super-cell composed of 3 x 3 x 3 unit cells. (A) Perspective along c-axis; (B) Orthographic along c-axis; (C) Perspective along a-axis, (D) Orthographic along a-axis. Holes are seen to cluster at the vacancy in proximal $O2p_\pi$ and $Cu3d_{xx-yy}$ orbitals.

**Figure 4.**

(A) Spin density in super-cell composed of 2 x 2 x 2 unit cells model of $Nd_{1.875}Ce_{0.125}CuO_4$ composed of $La_{14}NdCeCu_8O_{32}$. (B) Integrated DOS on Nd exhibiting $4f^3$ for the experimental crystal structure. (C) Same as (B) but for a 0.08 Å parallel displacement of the Nd-Ce pair relative to the $CuO_2$ plane renders $Nd4f^4$. (D) Same as top panel in Figure 2. (E) Three Nd-Ce $4f^4$ related electron configurations replace the super-atom states in Hg-1201 (lower panels in Figure 2). Emphasized are the symmetry properties of the Ce-Nd 4f orbitals, when projected onto the ab-plane.

**Figure 1.**

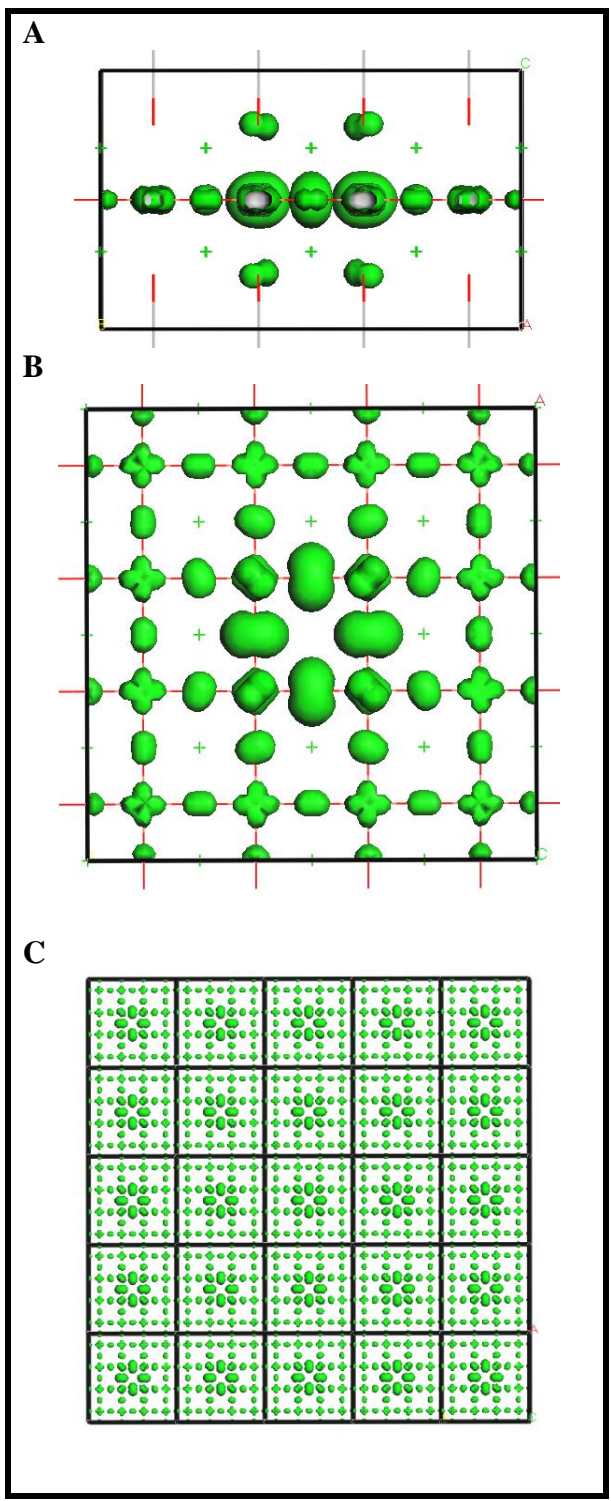

**Figure 2**

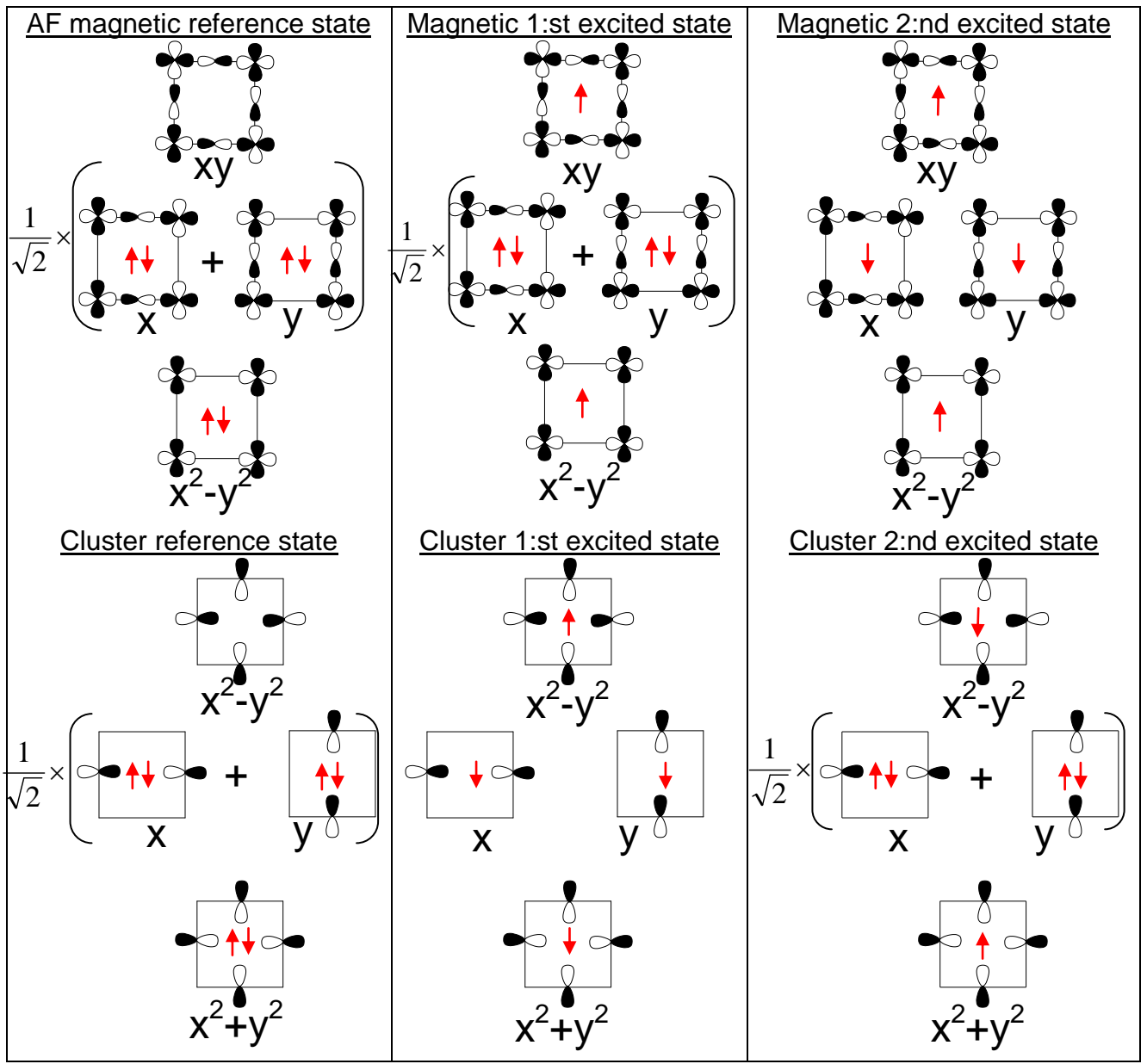

**Figure 3**

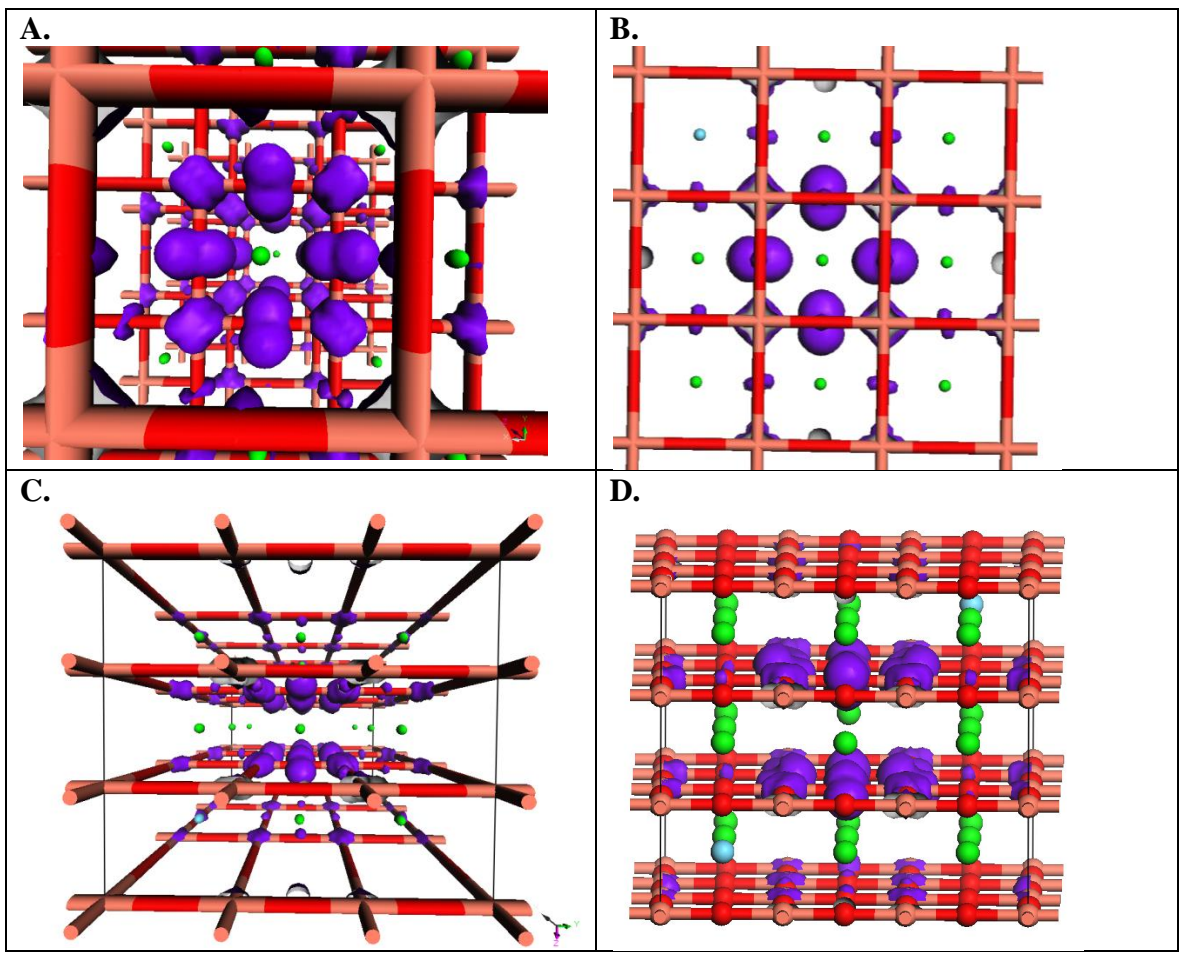

**Figure 4.**

A. B. C.

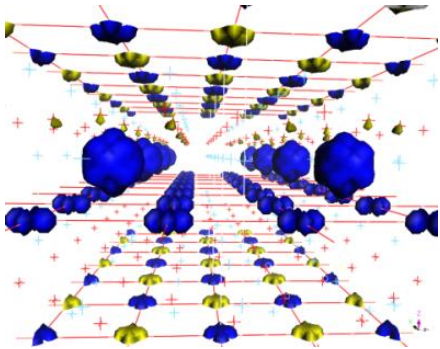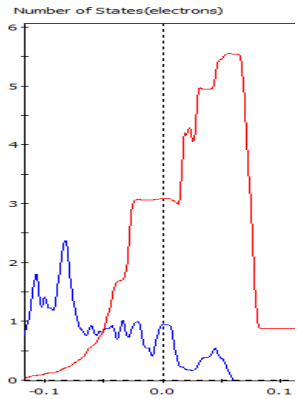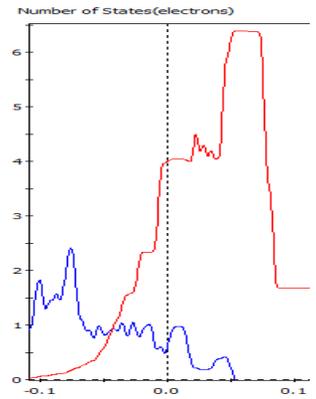

D.
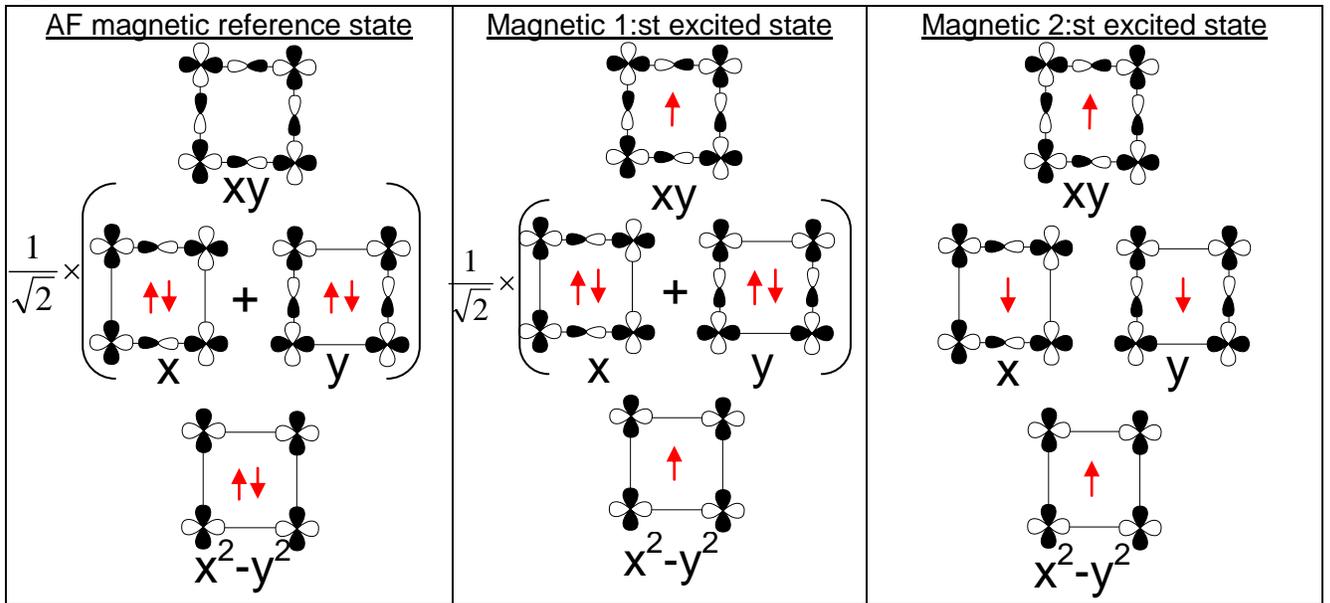

E.
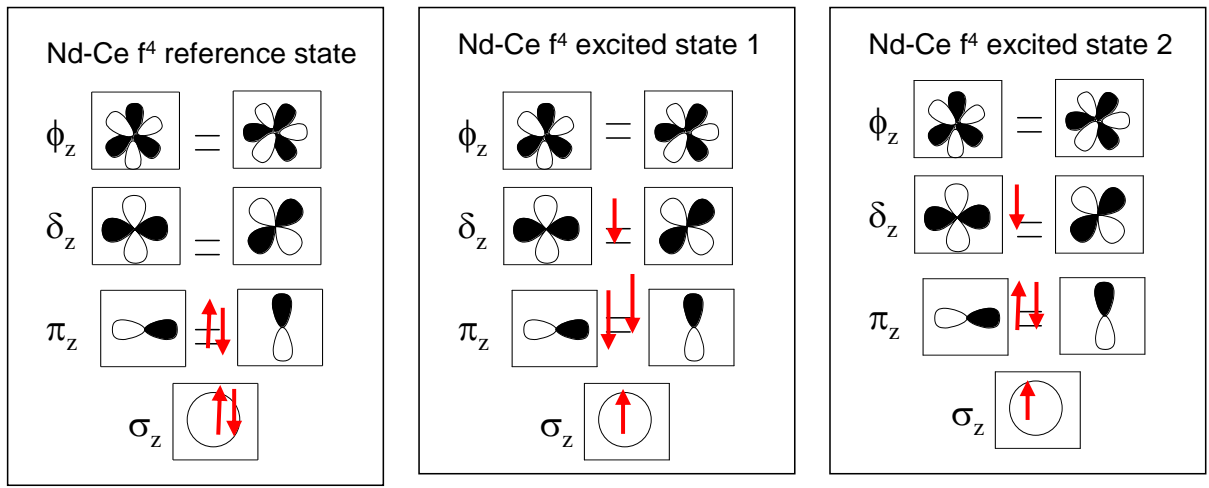